\documentclass[pra,twocolumn,floatfix,showpacs,superscriptaddress]{revtex4}
\usepackage{amsmath,amssymb,amsfonts,bbm,graphicx,times,psfrag}
\usepackage{amsthm}
\usepackage{bbm}
\usepackage{color}
\newcommand{\Tr}{{\rm Tr}}
\newcommand{\bra}[1]{\left\langle #1 \right\vert}
\newcommand{\ket}[1]{\left\vert #1 \right\rangle}
\newcommand{\daga}{^{\dagger}}

\newcommand{\TRAPrev}[1]{{#1}}
\newcommand{\JT}{\TRAPrev}

\def\sinc{\mbox{sinc}}
\def\sid{{\scriptscriptstyle C}}
\def\sE{{\scriptscriptstyle E}}
\def\sq{{\scriptscriptstyle Q}}
\def\sch{{\scriptscriptstyle C\!O\!H}}
\def\sho{{\scriptscriptstyle P\!H\!N}}
\begin{document}
\title{Quantum phase communication channels in the presence 
of static and dynamical phase diffusion}
\author{Jacopo Trapani}\email{jacopo.trapani@unimi.it}
\affiliation{Dipartimento di Fisica, Universit\`a degli Studi di 
Milano, I-20133 Milano, Italy}
\author{Berihu Teklu}\email{berihut@gmail.com}
\affiliation{
Institut Pascal, PHOTON-N2, Clermont Universit\'e, Blaise Pascal
University, CNRS, F-63177 Aubi\`ere Cedex, France.}
\author{Stefano  Olivares}\email{stefano.olivares@fisica.unimi.it}
\affiliation{Dipartimento di Fisica, Universit\`a degli Studi di 
Milano, I-20133 Milano, Italy}
\affiliation{CNISM, UdR Milano Statale, I-20133 Milano, Italy}
\affiliation{INFN, Sezione di Milano, I-20133 Milano, Italy}
\author{Matteo G.~A.~Paris}\email{matteo.paris@fisica.unimi.it}
\affiliation{Dipartimento di Fisica, Universit\`a degli Studi di 
Milano, I-20133 Milano, Italy}
\affiliation{CNISM, UdR Milano Statale, I-20133 Milano, Italy}
\affiliation{INFN, Sezione di Milano, I-20133 Milano, Italy}
\date{\today}
\begin{abstract}
We address quantum communication channels based on phase modulation of
coherent states and analyze in details the effects of static and 
dynamical (stochastic) phase diffusion. We evaluate mutual information 
for an ideal phase receiver and for a covariant phase-space-based 
receiver, and compare their performances by varying the number of 
symbols in the alphabet and/or the overall energy of the channel. 
Our results show that phase
communication channels are generally robust against phase noise,
especially for large alphabets in the low energy regime. In the presence
of dynamical (non-Markovian) noise the mutual information is preserved
by the time correlation of the environment, and when the noise spectra
is detuned with respect to the information carrier, revivals of mutual
information appears.
\end{abstract}
\maketitle
\section{Introduction}\label{s:intro}
The transmission of classical information along an ideal 
bosonic quantum channel is optimized by encoding information 
onto Fock number states, according to a thermal distribution, and 
then retrieving this information by the measurement of 
the number of photons \cite{yue93,dru94,hol01}. This strategy
allows to achieve the ultimate channel capacity, 
i.e. to maximize the mutual information between the sender 
and the receiver, given a constraint on the overall energy
sent through the channel, thus outperforming other encoding/decoding 
scheme involving different degrees of freedom of the radiation 
field, e.g. the amplitude or the phase.
\par
If we take into account the unavoidable noise affecting 
the information carriers along the channel, the situation 
becomes more involved and a question arises on whether different
coding/encoding schemes may offer better or comparable 
performances. 
\JT{Indeed, in the presence of a 
phase insensitive noise, e.g. amplitude damping, 
also coherent coding has been shown to 
achieve the ultimate channel capacity \cite{gio04,guh05}.
}
\par
In this paper, we address communication channels based on 
phase encoding \cite{fed81,ste96,hal93}
and analyze in details their performances in the
presence of phase diffusion, which represents 
the most detrimental kind of noise affecting this kind of channel
\cite{oli13, ph:diff:geno}.  In particular, we will consider 
communication schemes where the information is encoded by 
modulating the phase of a coherent signal, which then travels 
through a phase-diffusing environment before arriving at 
the receiver station and being detected. We consider two different 
environment models in which phase noise is either induced by a 
stationary environment inducing a static noise, or by a fluctuating
one leading to stochastic phase diffusion. We then evaluate the 
mutual information for both an ideal phase receivers and 
a covariant phase-space-based one (corresponding to the 
marginal phase distribution of the Husimi Q-function). We then
compare their performances each other and with the capacity of 
other relevant channels, including the optimal one.
Our results show that phase-keyed communication channels are 
robust against phase diffusion and offer performances comparable 
to channels involving coherent encoding. 
Phase channels may even approach the ultimate capacity in the 
low energy regime and for large alphabets.
\par
The paper is structured as follows. In Section 
\ref{sec:2} we describe the communication protocol 
details and derive a general formula for
the corresponding mutual information. In Section \ref{sec:3} we 
introduce a model for the static noise case and discuss 
the effects on the channel performance, making a comparison 
with the cases of photon number and amplitude channels. 
Section \ref{sec:4} considers channels affected by dynamical
(stochastic) phase diffusion and discuss the significative 
differences with the static case. Section \ref{s:out} closes
the paper with some concluding remarks.
\section{Phase-keyed communication channels\label{sec:2}}
\begin{figure}[h!]
\begin{center}
\includegraphics[width=0.45\textwidth]{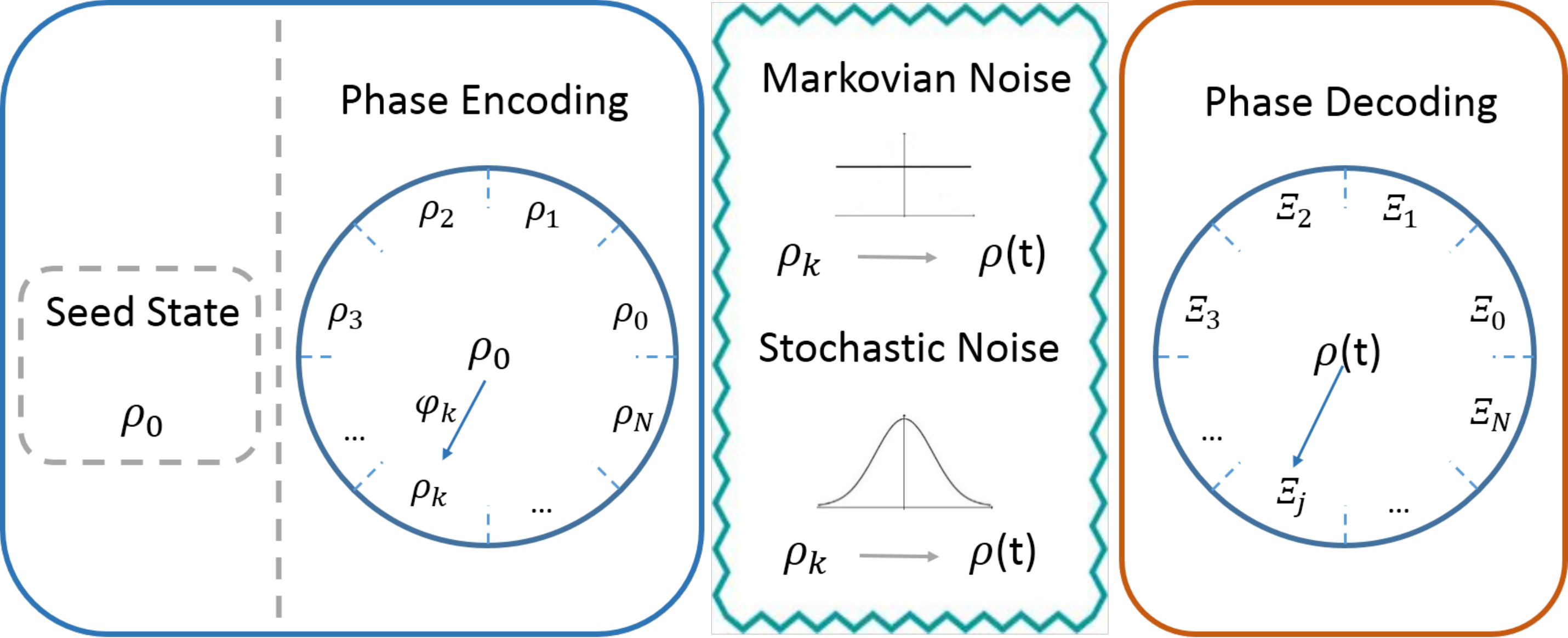}
\end{center}
 \caption{\label{f:scheme}
(Color online) Schematic representation of a phase 
communication channel. The sender encodes a finite
number $N$ of symbols using $N$ different values 
of a phase-shift $\phi_k = 2\pi k/N$ 
imposed to a {\it seed} coherent state $\rho_0$.
The signal then propagates along the channel, 
to the  receiver station, in the
presence of either static or dynamical noise
and it is finally detected by a suitable 
measurement scheme in order to retrieve 
the carried information. }
 \end{figure}
\par
A schematic diagram of a quantum phase communication channel 
is depicted in Fig. \ref{f:scheme}.
The sender encodes a finite number $N$ of symbols using 
$N$ different values of a phase shift $\phi_k$, where 
$\phi_k < \phi_j$ if $k< j$ and $0\le k < N$.
We assume a choice of equidistant phase values 
$\phi_k = 2\pi k/N$. The phase $\phi_k$ is encoded 
onto a {\it seed} state $\rho_0$ of a single-mode 
radiation field by the unitary phase-shift operation 
$U(\phi)=\exp(i \phi\,  
a^{\dagger} a )$, $a$ being the annihilation operator,
$[a,a^{\dag}]=1$, namely:
\begin{equation}
\label{eq:encoding}
\rho_0 \to \rho_k \equiv
U(\phi_k) \rho_0 U^{\dag}(\phi_k)\,.
\end{equation}
The signal then propagates along the channel  to the 
receiver station, where it is detected by a suitable 
measurement scheme in order to retrieve the information it 
carries. More explicitly: the receiver performs a phase 
measurement on the output state and, once the phase is measured, 
\TRAPrev{she} chooses a strategy to associate the measured value to one 
of the symbols of the sender's alphabet.
The inference strategy should match the (continuous)
output from the phase measurement to a symbol from a discrete
alphabet. The straightforward choice consists in 
associating each phase value with the closest $\phi_k$ 
within a margin of error. To this aim 
the receiver divides the full phase range 
$[0,2 \pi)$ into $N$ bins, corresponding to 
the intervals $$\Xi_j = [\phi_j-\Delta, \phi_{j}+\Delta)\,,$$
where $\Delta=\pi/N$ and 
$\bigcup_{j=1}^{N} \Xi_j = [0,2 \pi)$.  
More generally, the width of each bin may be different and 
dependent on $j$, though a symmetric choice is 
often optimal and will be assumed throughout this paper.
If $\phi$ denotes the value of the receiver's outcome, 
we express the inference rule as follows:
\begin{equation}
\label{eq:strategy}
\mbox{if} \quad \phi \in \Xi_j \Rightarrow \quad \phi 
\rightarrow \phi_j.
\end{equation}
The positive operator-valued measure (POVM)
$\{\Pi(\phi_{j})\}\equiv \{ \Pi_{j}\}$ describing the
measurement strategy employed by the receiver can be written as:
\begin{equation}
\label{eq:generalPOVM}
\Pi_{j}=\int_{\phi_{j}-\Delta}^{\phi_{j}+\Delta}\pi(\theta)d\theta,
\end{equation}
where $\pi(\theta)$ is the actual POVM of the 
phase measurement performed by the receiver. A POVM
for a covariant phase measurement may always written 
as \cite{boh95,can96}:
\begin{equation}
\label{eq:POVM} 
 \pi(\theta)=\frac {1} {2\pi} \sum_{n,m=0}^{\infty} 
 A_{n,m} e^{-i (n-m)\theta} \ket n\bra m,
\end{equation}
where $A_{n,m}$ are the elements of a positive and Hermitian matrix
${\boldsymbol A}$, which is measurement-dependent. Covariance 
follows easily from Eq. (\ref{eq:POVM}), since 
$U(\phi) \pi (\theta) U\daga (\phi) = \pi (\theta+ \phi)$
and thus 
\begin{equation}\label{eq:covariant} 
 \Pi_{j}=
U(\phi_j) \Pi_{0} U^{\dag}(\phi_j).
 \end{equation}
The combination of  Eqs. ($\ref{eq:generalPOVM}$) and (\ref{eq:POVM}) 
brings to an explicit form of the POVM $\Pi_j$, given by
\begin{equation}
\label{eq:POVM2}
\Pi_j = \sum_{n,m=0}^{\infty} A_{n,m} f_{n-m}(j) |n \rangle \langle m|
\end{equation}
where the structure of the POVM is determined by the resolution function
\begin{equation}
\label{eq:resolution}
f_d(j) = \frac{1}{2 \pi} \int_{\phi_j -
\Delta}^{\phi_j+\Delta} e^{-i d \theta} d 
\theta = \frac{\sin \Delta \pi}{\pi d} e^{-i d \phi_j},
\end{equation}
with the property $\sum_{j=1}^{N} f_d (j) = \delta_{d,0}$,
 where $\delta$ is the Kronecker delta. 
\par
The figure of merit to assess the performances of a communication
channel is the mutual information between sender and receiver.
This quantity measures the amount of information shared by the two 
parties and can be written as
\begin{align}
\label{eq:mutINF}
 I&=\sum_{k,j=0}^{N-1} p(k,j)\log_2\frac{p(k,j)}{p(k)\,p'(j)} \nonumber\\
 &=\sum_{k,j=0}^{N-1} p(j|k)p(k)\log_2\frac{p(j|k)}{p'(j)},
\end{align}
where $p(j|k)$ is the conditional probability of measuring a phase $\phi_j$
given the input phase $\phi_k$; 
$p(k)$ is the {\it{a priori}} probability distribution of 
transmitting a $\phi_k$-encoded seed state;
$p(k,j) = p(j|k)\,p(k)$ is the joint probability to send the
symbol $\phi_k$ and obtaining the outcome $\phi_j$ and, finally, 
$p'(j)\equiv p'(\phi_j)$ is 
the probability of the outcome $\phi_j$, given by
%
$p'(j)=\sum_{k=0}^{N-1}p(j|k)p(k)$.
%
\par
Maximization over the probability $p(\phi_k)$ leads 
to the so called channel capacity, i.e.
the maximum information transmitted 
through the channel per use. 
In particular, we analyze the case of uniform 
encoding probability, $ p(k) = N^{-1}$, 
i.e. the letters have the
same probability to be sent through the channel.
The conditional probability of an outcome $\phi$ falling in the
bin $\Xi_{j}$ given the initial state $\rho_{k}$ is
\begin{equation}
\label{eq:condprob}
p(\phi\in\Xi_{j}|\rho_k) \equiv p(j|k)=\Tr[\rho_{k}\Pi_{j}].
\end{equation}
Under these conditions, the mutual information reduces to
\begin{equation}\label{eq:mutINF2}
I=\frac{1}{N}\,\sum_{k,j=0}^{N-1} \Tr[ \varrho_{k} \Pi_{j}]
 \log_2\left\{\frac{\Tr[ \varrho_{k} \Pi_{j}]}
{N^{-1}\sum_{h=0}^{N-1}\Tr[ \varrho_{h} \Pi_{j}]}\right\}.
\end{equation}
\par
By using the covariance property of the POVM and its explicit form given 
in Eq. (\ref{eq:POVM2}), the conditional probability can be expressed as
\begin{align}
\label{eq:prob}
  p(j|k) &=\Tr[\rho_{k}\Pi_{j}]  = \Tr[\rho_{0}\Pi_{j-k}] \nonumber \\ 
 &=  \sum_{n,m=0}^{\infty} A_{n,m} f_{n-m}(j-k) \rho_{n,m}.
\end{align}
\JT{Note that $\sum_{k} p(j|k) = \sum_k \mbox{Tr}[\rho_k\Pi_j] = 1$, which follows
from the symmetries of the resolution function, $f_{-d}(j) = f_d (j)$, i.e. $f_{-d}(-j) = f_d(j)$.  }
Upon introducing the positive quantity $s = |j-k|$, we
obtain a simpler form for the mutual information
\begin{equation}
\label{eq:mutINFq}
I \equiv I(N, \bar n) = \log_2 N + \sum_{s=0}^{N-1} q(s) \log_2 q(s)
\end{equation}
where $\bar n$ is the average number of photons of the seed signal and
\begin{flalign}\label{eq:q}
q(s)&=\sum_{n,m=0}^{\infty}A_{n,m}f_{n-m}(s)\rho_{n,m} \\
&= \frac{1}{N}\Big\{1 + \sum_{n=0}^{\infty}\sum_{d=1}^{\infty} A_{n,n+d}\left[ f_d (s) \rho_{n,n+d}+c.c. \right] \Big\}.
\end{flalign}
The function $q(s)$ measures the probability of finding a $2 \pi s/N$
phase difference between the input and output signal, whatsoever value
the encoded phase may assume.
\par
The function $q(s)$ and thus the performances of the communication
channel do depend on the measurement performed by the receiver through
the matrix $(A_{n,m})$ and on the seed state via the matrix elements 
$\rho_{n,m} = \langle n| \rho_0| m \rangle$.
In the following, we will focus on two particular phase 
measurements: the canonical phase measurement \cite{lyn95,boh95,can96,
roy96,par99} and a phase-space-based one, i.e. the marginal 
phase distribution obtained from the Husimi $Q$-function 
\cite{ man92,hra93,noh93,vog93,
dar93,ulf93,dar94,wis95,ops95,pel11}. The latter is a feasible
phase measurement achievable, e.g., by heterodyne or double-homodyne
detection. For the canonical measurement we have $A_{n,m}=1$, whereas for 
the $Q$-measurement $A_{n,m} = \Gamma[1+\frac{1}{2}(n+m)](n!m!)
^{-\frac{1}{2}}$, $\Gamma[x]$ being the Euler Gamma function.
\section{Quantum phase communication
channels in the presence of static phase diffusion}\label{sec:3}
In this section we address quantum phase 
communication channels in the presence of phase diffusion, and start
by considering situations where the environmental noise is static.
Any phase communication channel is based on the observation that
the optical field produced by a laser provides a convenient quantum system 
for carrying information. 
In particular, coherence of laser source ensures that a well-defined 
phase can be attributed to a light mode. Still, the unavoidable presence 
of noise generates a phase diffusion, which ultimately limits the 
coherence of the light.  
The master equation governing the evolution of the light beam 
in a static phase diffusing environment may be written as 
\cite{ph:diff:geno,ph:diff:geno:2}:
\begin{equation}
\label{eq:meq}
\frac {d} {dt}\rho = \frac{\Gamma}{2}
\mathcal{L}[a^{\dag} a] \rho\,,
\end{equation}
where $\mathcal{L}[O]\rho = 2 O \rho O^\dag - O^\dag O
\rho - \rho O^\dag O$ and $\Gamma$ is the static phase noise
factor.  An initial state $\rho(0)$ evolves with time as
\begin{equation}
\label{phase:noise}
 \rho(t)= \sum_{n,m=0}^{\infty}e^{- 
 \frac{1}{2} \tau(n-m)^2 }  \rho_{n,m} |n\rangle\langle m|\,,
\end{equation}
where we introduced the rescale time $\tau = \Gamma t$.
One can easily see that the diagonal elements $\rho_{n,n}$
are unaffected by the phase noise, thus, energy is conserved, whereas
the off-diagonal elements decay away exponentially. 
\par
In the rest of our paper we assume that the input seed is a coherent
state of the radiation field, namely, $\rho_0 =
\ket{\alpha}\bra{\alpha}$ with:
\begin{equation}
\label{eq:cohstate}
\ket\alpha=e^{-|\alpha|^2/2}\sum_{n=0}^{\infty}\frac
{\alpha^{n}}{\sqrt{n!}}\ket n.
\end{equation}
 Without lack of
generality, we assume $\alpha$ to be real. The
density matrix elements associated with the initial coherent state
$\rho_0$ are
\begin{align}
\label{eq:CS}
\rho_{n,m}=e^{-\bar n} \frac{\bar n^{(n+m)/2}} {\sqrt{n!m!}},
\end{align}
where $\bar n \equiv \alpha^{2}$
is the average number of photons of the coherent state $\rho_0$.
Exploiting Eq.~(\ref{phase:noise}), we find that the state arriving at
the receiver after the propagation through the noisy channel has the
following density matrix elements:
\begin{align}
\label{eq:CStime}
\rho_{n,m} \to \rho_{n,m}(t)=
 \, e^{- \frac{1}{2} \tau (n-m)^2}\rho_{n,m},
\end{align}
which can be used to evaluate the mutual information as written in
Eq. (\ref{eq:mutINFq}) once the POVM describing the measurement is given
and, thus, the $f_{n-m}(s)$ are assigned.
\par
The POVM describing the ideal (canonical) measurement is obtained from
Eq.~(\ref{eq:POVM}) with $A_{n,m} = 1$, $\forall n,m$.
In turn, the probability $q(s)$ after the phase diffusion
process reads:
\begin{align}
\label{eq:qID}
q_\sid
(s)=&\frac {1}{N}\biggl\{1+2e^{ -\bar{n}}\sum_{n=0}^{\infty}
\sum_{d=1}^{\infty} \sinc \left({\frac {\pi d} {N}}\right)\, 
e^{-\frac12 d^2 \tau} \notag\\
&
\times\cos \left[\frac{\pi d}{N}(2s+1)\right]
\frac {\bar{n}^{n+d/2}} {\sqrt{n!(n+d)!}}\biggr\},
\end{align}
where $\sinc(x) = \sin(x)/x$. The quantum mutual
information $I_{\sid}$ directly follows from 
Eq.~(\ref{eq:mutINFq}).
\begin{figure}[tb]
\vspace{0.5cm}
\begin{center}
\includegraphics[width=0.45\columnwidth]{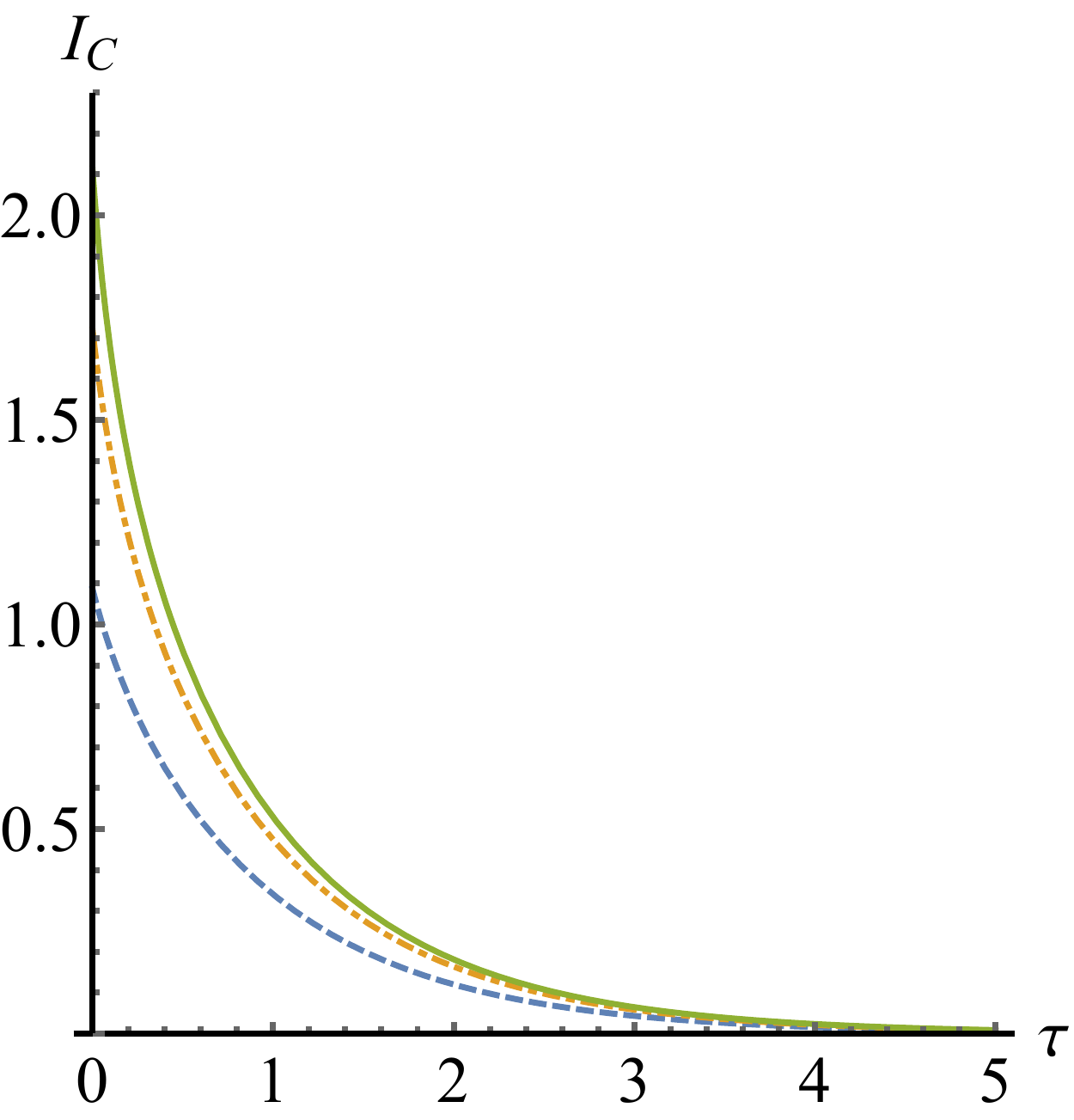}
\includegraphics[width=0.45\columnwidth]{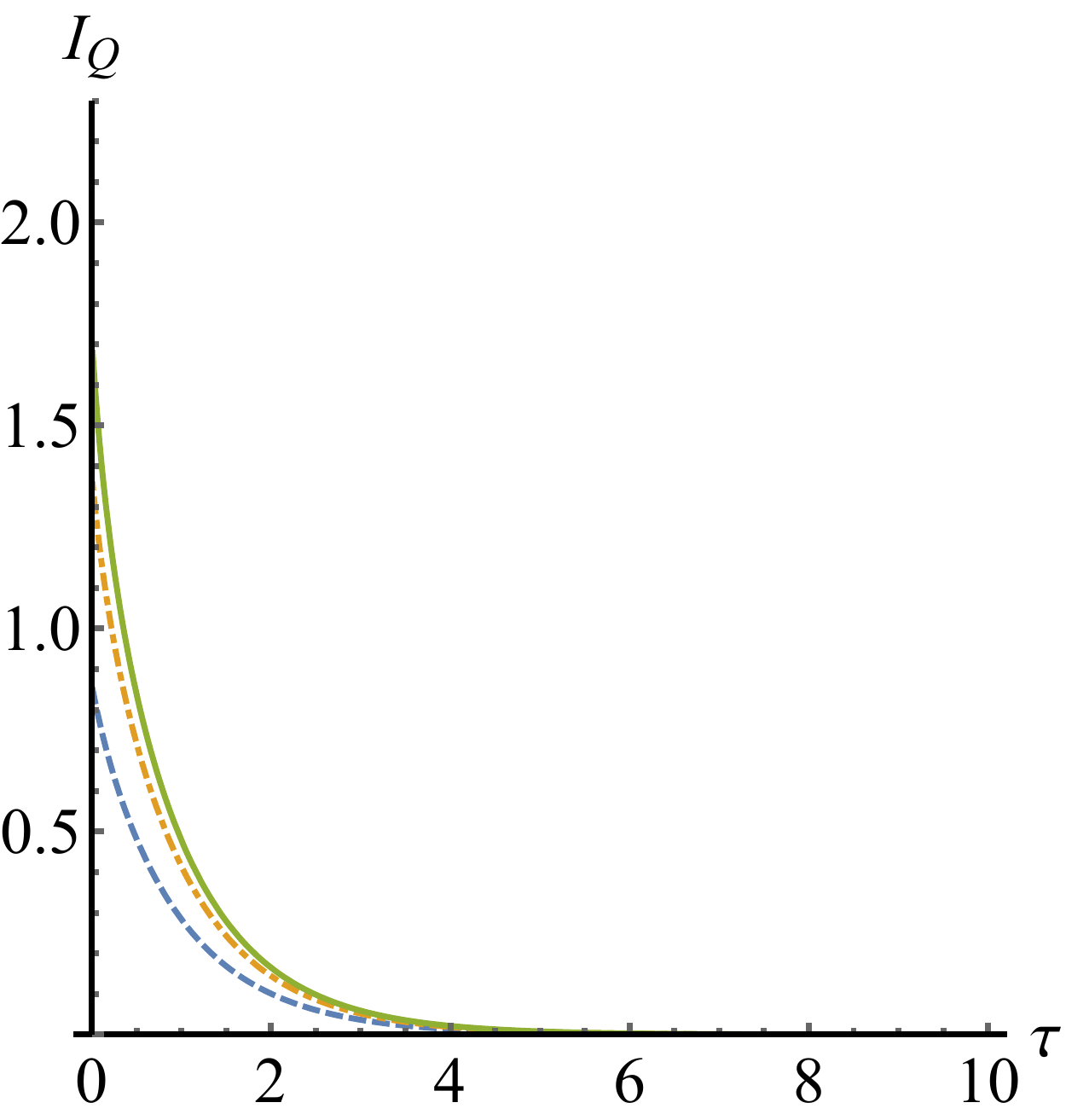}\\
\vspace{0.5cm}
\includegraphics[width=0.7\columnwidth]{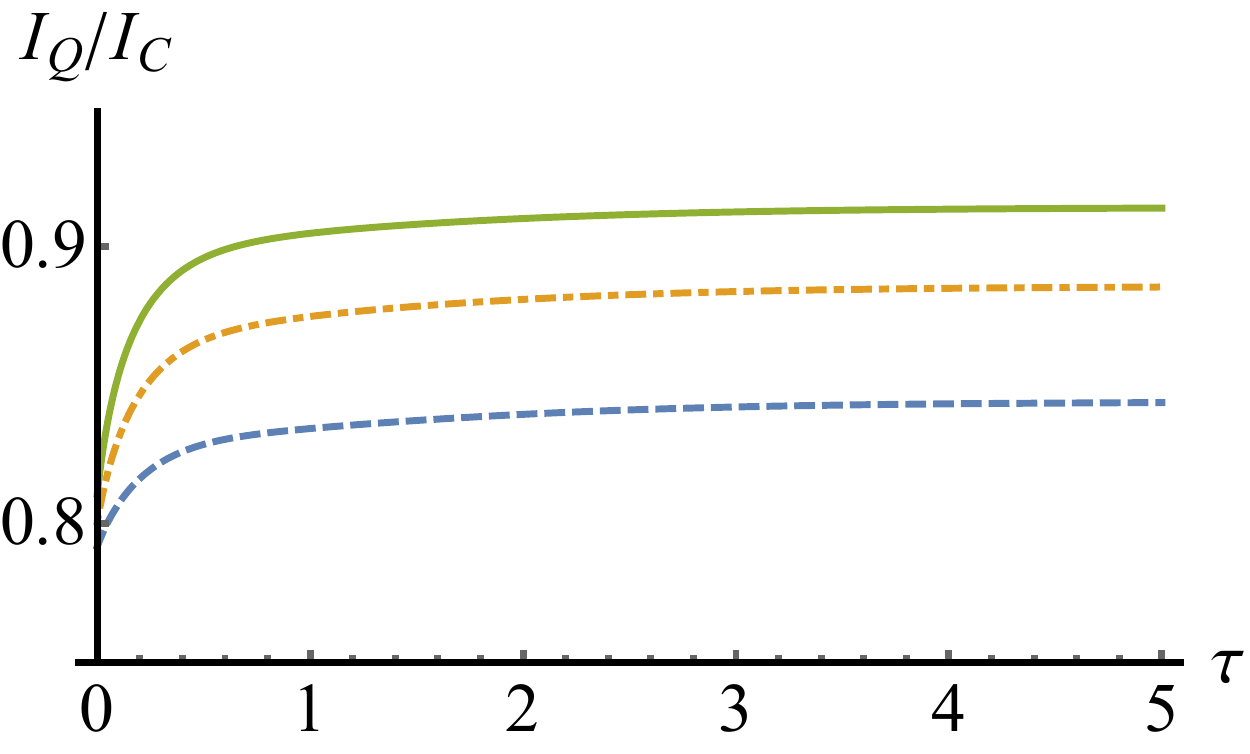}
\end{center}
\vspace{-0.5cm}
 \caption{\label{f:mutInf1}
(Color online) Phase communication channels in the presence of static
phase diffusion. The upper panels show the mutual informations 
for the ideal receiveir $I_\sid$ (left) and the $Q$ one $I_\sq$ (right) 
as a function of the noise parameter $\tau = \Gamma t$ 
for different
values of the average number of photons: from bottom to top, 
$\bar n=1$ (dashed blue), $\bar n= 2$ (dotdashed orange), $\bar n =3$ 
(solid green). We set the alphabet size to $N=20$.
The lower panel shows the ratio $I_\sq/I_\sid$ 
as a function of $\tau$ for different values of the average 
number of photons: from bottom to top, $\bar n=1$ (dashed blue), $\bar
n=2$ (dotdashed orange), $\bar n =3$ (solid green).} 
\end{figure}
\par
The probability $q_\sq(s)$ for the $Q$-measurement process
is obtained using 
$A_{n,m} = \Gamma[1+\frac{1}{2}(n+m)](n!m!)^{-\frac{1}{2}}$:
\begin{align}\label{eq:MUQ}
q_\sq(s)=&\frac {1}{N}\biggl\{1+2e^{ -\bar{n}}\sum_{n=0}^{\infty}
\sum_{d=1}^{\infty} \sinc \left({\frac {\pi d} {N}}\right) 
e^{-\frac12 d^2 \tau}\notag\\
&
\times \cos \biggl[\frac{\pi d}{N}(2s+1)\biggr]
\frac {\Gamma(1+n+\frac{d}{2})\bar{n}^{n+d/2}} {n!(n+d)!}\biggr\}.\notag\\
\end{align}
The corresponding mutual information $I_\sq$ is again obtained using
Eq.~(\ref{eq:mutINFq}).  
\par
In the upper panels of Fig. \ref{f:mutInf1} we show the mutual
information as a function of the rescaled time variable 
$\tau$, which plays the role of a noise parameter, 
for ideal (upper left panel) and $Q$ (upper right panel) 
phase-receivers and for different values $\bar n$ of the 
average number of photons of the seed state. The size of the 
alphabet is set to $N=20$.
As it is apparent from the plots, phase diffusion 
leads to an unavoidable  loss of information.
The mutual information $I_\sq$ for $Q$ receivers shows 
the same vanishing behavior in time as the 
ideal one  $I_\sid$, though its value is always slightly 
smaller. In order to provide a quantitative assessment 
we show their ratio $I_\sq/I_\sid$ in the lower panel of the 
same figure, as a function of $\tau$ 
for different values of $\bar n$. The ratio is always below 
one, thus confirming that $Q$ receivers are not as 
efficient as the ideal ones. The ratio slighty 
increases with time, i.e. for long distance channels, and with 
the energy of the seed signal.
\par
In order to further assess the performances of phase 
channels we now 
compare the mutual informations $I_\sid$ and $I_\sq$ with the 
capacity of a (realistic) coherent channel and with the ultimate quantum
capacity of a single-mode channel, which is achieved by the photon number
channel. In a coherent channel information is encoded onto the 
amplitude of a coherent signal and then retrieved by heterodyne or
double-homodyne detectors, the channel capacity is achieved 
by Gaussian modulation of the amplitude and is given by 
\begin{equation}
\label{eq:realistic}
C_{\sch}(\eta) = \log (1+ \eta \bar n)\,,
\end{equation}
where $\bar n$ is again the average number of photon per use of 
the channel, and  $\eta$ is the overall (amplitude) loss along 
the channel. On the other hand, the ultimate quantum capacity of 
a single-mode channel, which also saturate\TRAPrev{s} the Holevo-Ozawa-Yuen 
bound \cite{yue93}, is achieved by the photon number channel
\begin{equation}
\label{eq:holevo}
C_{\sho} = (\bar n +1) \log_2 (\bar n+1) - \bar n \log_2 \bar n.
\end{equation}
where information is encoded onto the number
of quanta transmitted through the channel according to a thermal
distribution, and the decoding stage is performed by 
photodetection. 
\par
\begin{figure}[tb]
\vspace{0.5cm}
\begin{center}
\includegraphics[width=0.45\columnwidth]{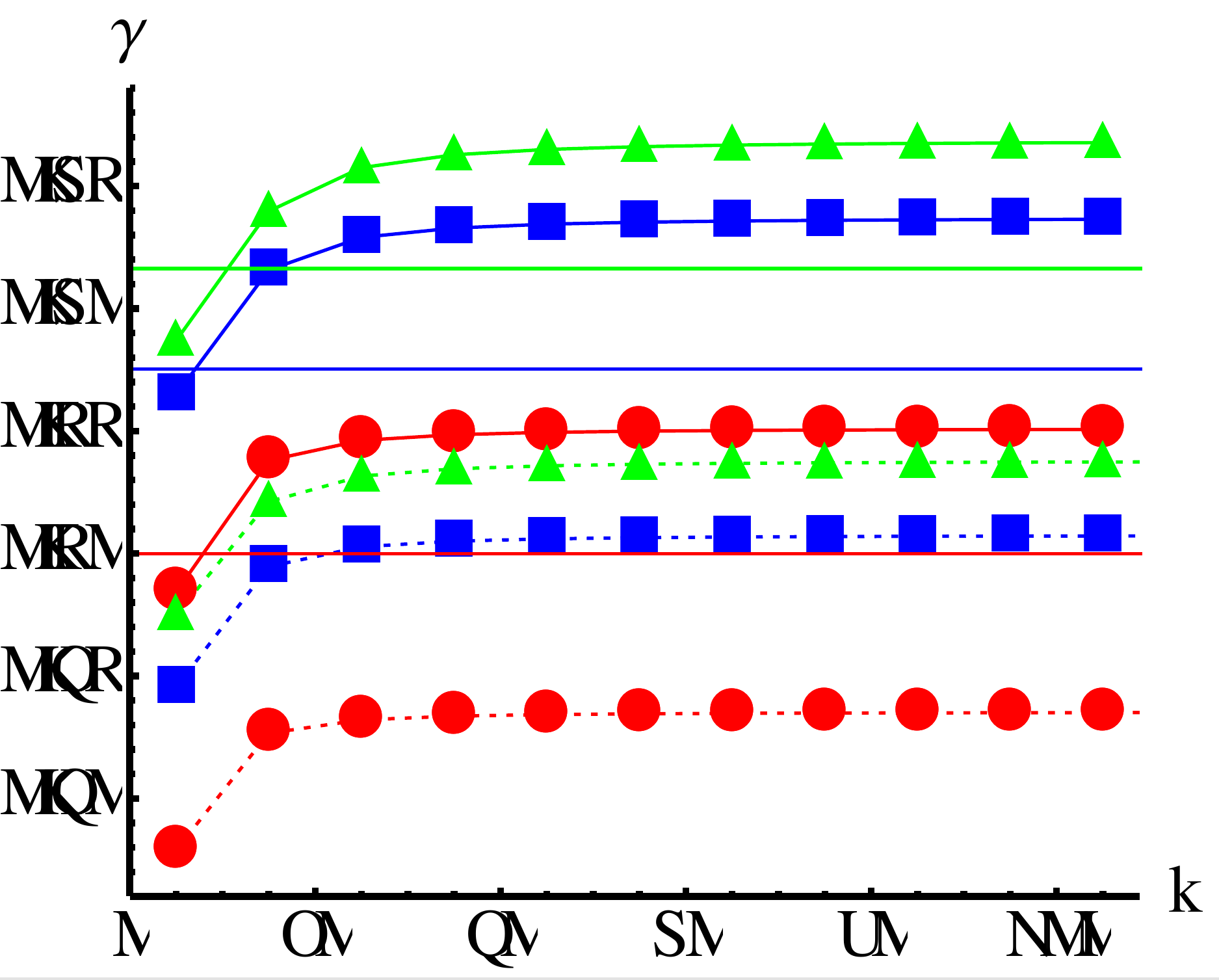}
\includegraphics[width=0.45\columnwidth]{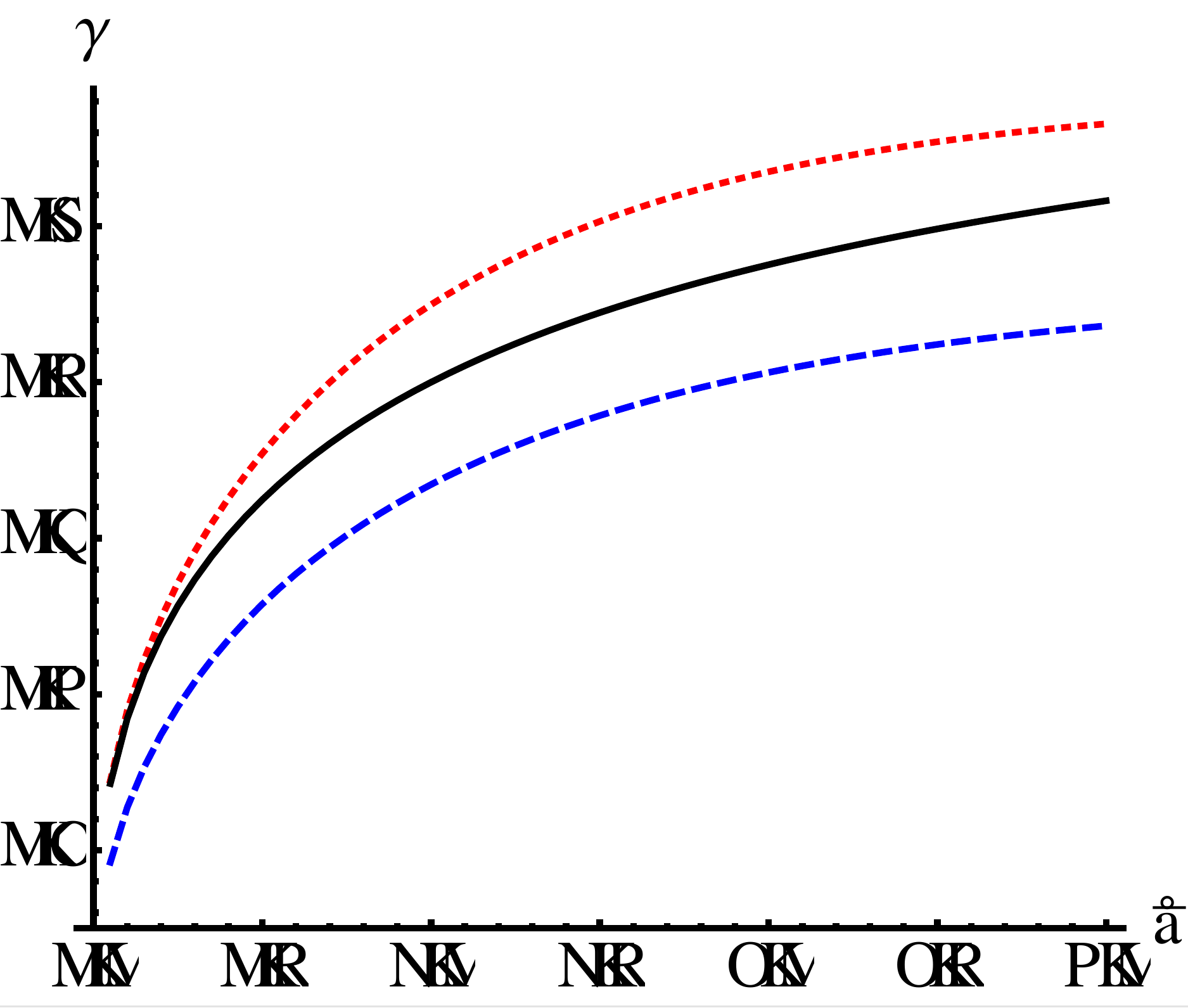}\\
\vspace{0.5cm}
\includegraphics[width=0.45\columnwidth]{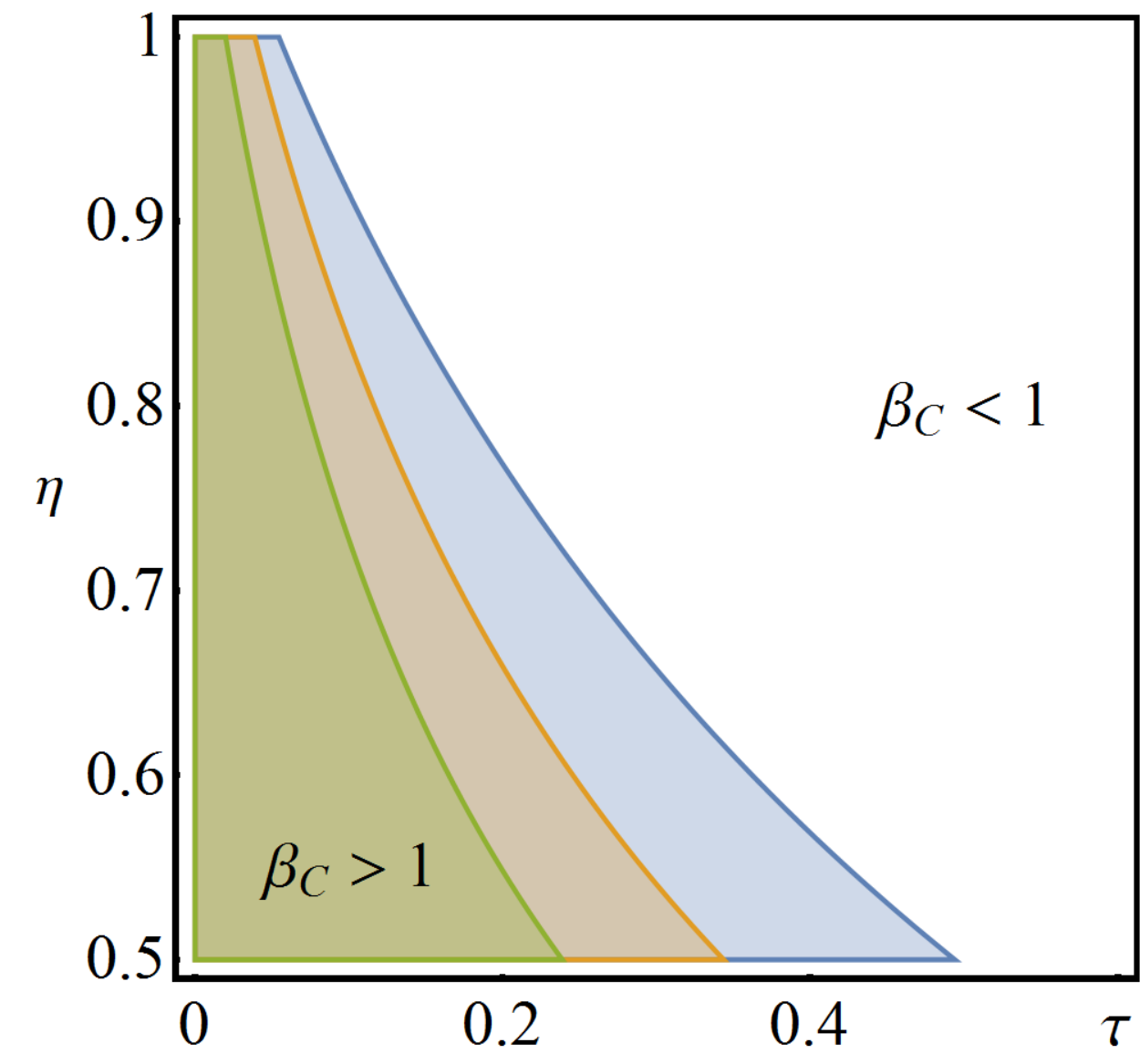}
\includegraphics[width=0.45\columnwidth]{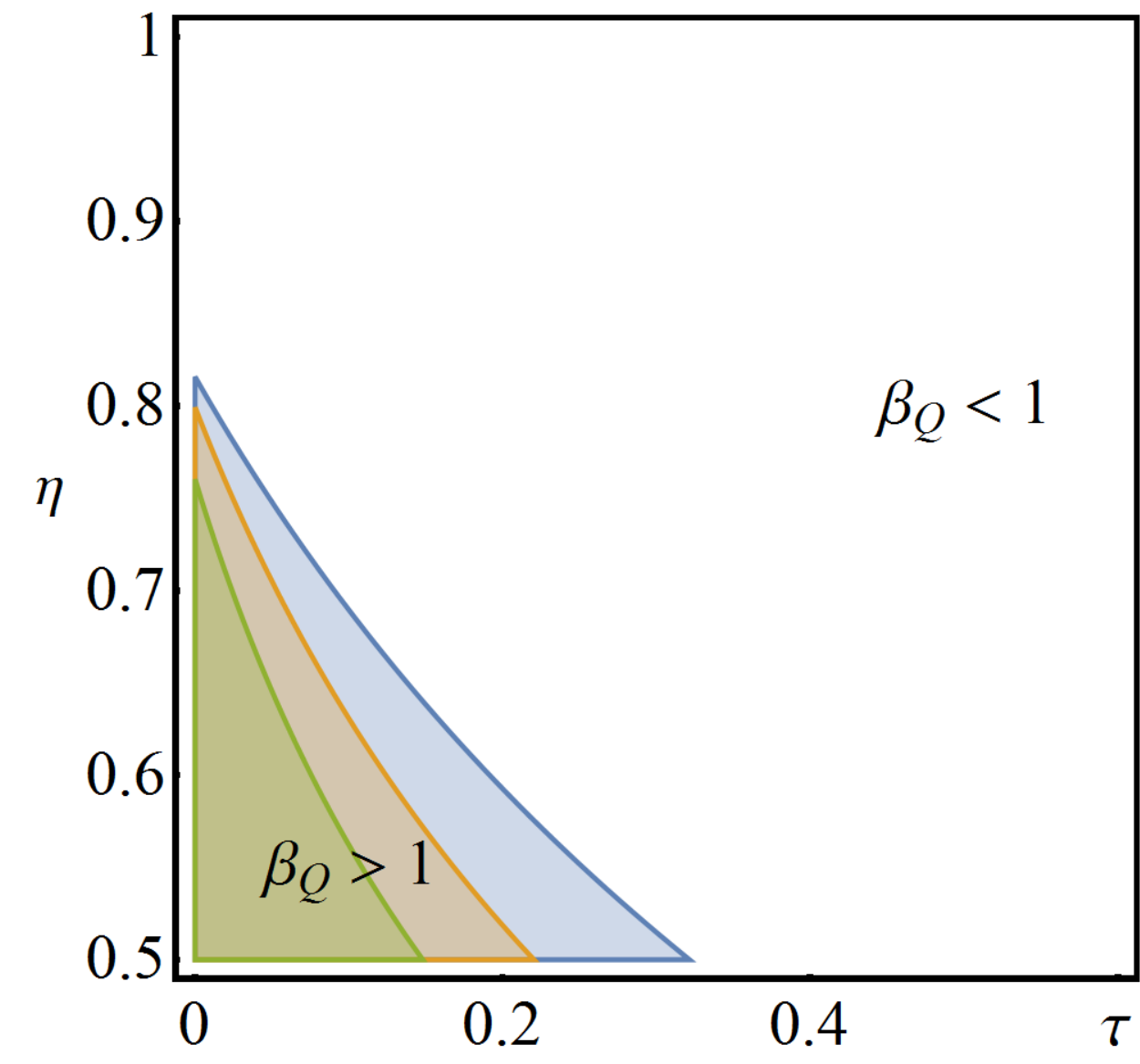}
\end{center}
\vspace{-0.5cm}
 \caption{\label{f:perfs} 
 (Color online) The upper left panel shows the ratios $\gamma_\sid$
(symbols joined by solid lines)
and $\gamma_\sq$ (symbols joined by dotted lines) as a function of the
number of symbols in the alphabet for noiseless phase channels ($\eta = 1$).
Red  circles correspond to $\bar n=1$, 
blue squares to $\bar n=2$ and green triangles to $\bar n=3$. Solid
lines are the ratios $C_\sch/C_\sho$ for the same three values of $\bar n$ 
(from bottom to top) with the same color code. The upper right panel
shows the ratios $\gamma_\sid$ (dotted red), $C_\sch/C_\sho$ 
(solid black) and $\gamma_\sq$ (dashed blue) for noiseless channels 
as a function of $\bar n$ and for a fixed value of $N=50$. 
\JT{The lower panels show the regions $\beta_\sid>1$ and $\beta_\sq>1$,
respectively, as functions of $\tau = \Gamma t$ and $\eta$.
From left to right we have the regions corresponding
to  $\bar n=1,2,3$  (green, orange and blue, respectively). When 
$\beta_k >1$, $k=C,Q$, the phase channels become more effective than 
coherent ones. The boundary of each region singles out an
energy-dependent threshold on the noise parameters.}
} \end{figure}

At first, let us address noiseless phase channels and consider, 
for both receivers, the ratio between the corresponding mutual
information and the ultimate capacity, i.e. $\gamma_\sid = 
I_\sid/C_\sho$ and $\gamma_\sq = I_\sq/C_\sho$. The two quantities 
are reported in the upper left panel of Fig. \ref{f:perfs} as a function 
of the number of symbols in the phase alphabet, and for different 
values of the average number of photons $\bar n$. The plots reveal 
that an alphabet of about $N\simeq 50$ symbols is enough \JT{to reach
the asymptotic value} of both ratios, and in turn of $I_\sid$ and
$I_\sq$. Also, the plots show that the ratio with the ultimate capacity
is comparable to that of noiseless coherent channels, with ideal phase
receivers slightly outperforming
the coherent channel and the $Q$ one being slightly outperformed.
Using this size of the alphabet, we have evaluated $\gamma_\sid$ and
$\gamma_\sq$ as a function of the average photon number $\bar n$.
Results are shown in the upper right panel of Fig. \ref{f:perfs}, 
confirming that phase channels with ideal receivers performs 
slightly better than coherent ones, whereas $Q$ receivers lead 
to slightly worse performances.
\par
Let us now compare phase channel with coherent ones in the presence 
of noise. In the lower panels of Fig. \ref{f:perfs} we show the 
ratios $\beta_k = I_k/C_\sch$, $k=$ $C,Q$ between the mutual
information of our phase channels and the capacity of the coherent
channel as a function of the noise parameters, $\tau$ and $\eta$ of 
the two channels. Results for different values of the average 
number of photons $\bar n$ are shown. In both cases an 
energy-dependent threshold on the amount of noise appears, above which 
phase channels become more effective than coherent ones.
\par
Finally, let us discuss the performances of the two receivers in the
relevant quantum regime of low number of photons, $\bar n \ll 1$, and
large number of letters, $N \gg 1$. As it can be argued from the 
upper right of Fig. \ref{f:perfs}, both $I_\sid$ and $I_\sq$ grow 
linearly with $\bar n$ for $\bar n \ll1$, and this resembles the
behaviour of both the coherent capacity and the ultimate quantum
capacity. This means that, albeit being suboptimal, phase channels 
offer good performances when low energy should be transmitted through 
the channel.  This
finding can be confirmed by expanding the mutual information up to
the first order in the average photon number of the seed signal,
arriving at the expressions
\begin{equation} \label{eq:MIs}
I_{\rm ID}\,\stackrel{\bar n \ll 1}{\simeq}\,\frac{\bar{n}\, 
\sinc^{2}(\frac {\pi}{N})\, e^{-\tau}}{ \log 2}\,
\stackrel{N \gg 1}{\simeq}\,
\frac {\bar{n}\, 
e^{-\tau}}  {\log 2}
\end{equation}
for the ideal measurement and 
\begin{equation}\label{eq:MIqms}
I_{\rm Q}\,\stackrel{\bar n \ll 1}{\simeq}\,\frac{\pi}{4}\frac{\bar{n}\, 
\sinc^{2}(\frac {\pi}{N})\, e^{-\tau}}{ \log 2}\,
\stackrel{N \gg 1}{\simeq}\,\frac{\pi}{4}
\frac {\bar{n}\, 
e^{-\tau}}  {\log 2}
\end{equation}
for the $Q$-receiver one, their ratio approaching
the limiting value of $\pi/4$.
\section{Dynamical phase diffusion}
\label{sec:4}
In many experimental situations, the exchange of 
information between sender and receiver takes place in 
noisy environments which cannot be described in terms 
of a Markovian master equations.  In such cases, a
full quantum description of the interaction may be inconvenient, as the
approximations needed to obtain solvable dynamical equations 
could preclude the study of interesting features of the dynamics 
itself. On the other hand, when the exact quantum description is 
not achievable, it is still possible to model the interaction by 
classical stochastic fields (CSF), which happen to be
reliable models of quantum environments, especially when the noise
presents classical features, e.g. a Gaussian noise
\cite{cro14,ben14,cat15}.  
Moreover, the use
of a CSF also gives the chance to analyze in a simple way the role of
the correlation time of the environment, and the influence on the
dynamics of the presence of a detuning between the mode playing the role
of information carrier and the central (natural) frequency of the
environment.
\par
\begin{figure}[tb!]
\vspace{0.5cm}
\begin{center}
\includegraphics[width=0.45\columnwidth]{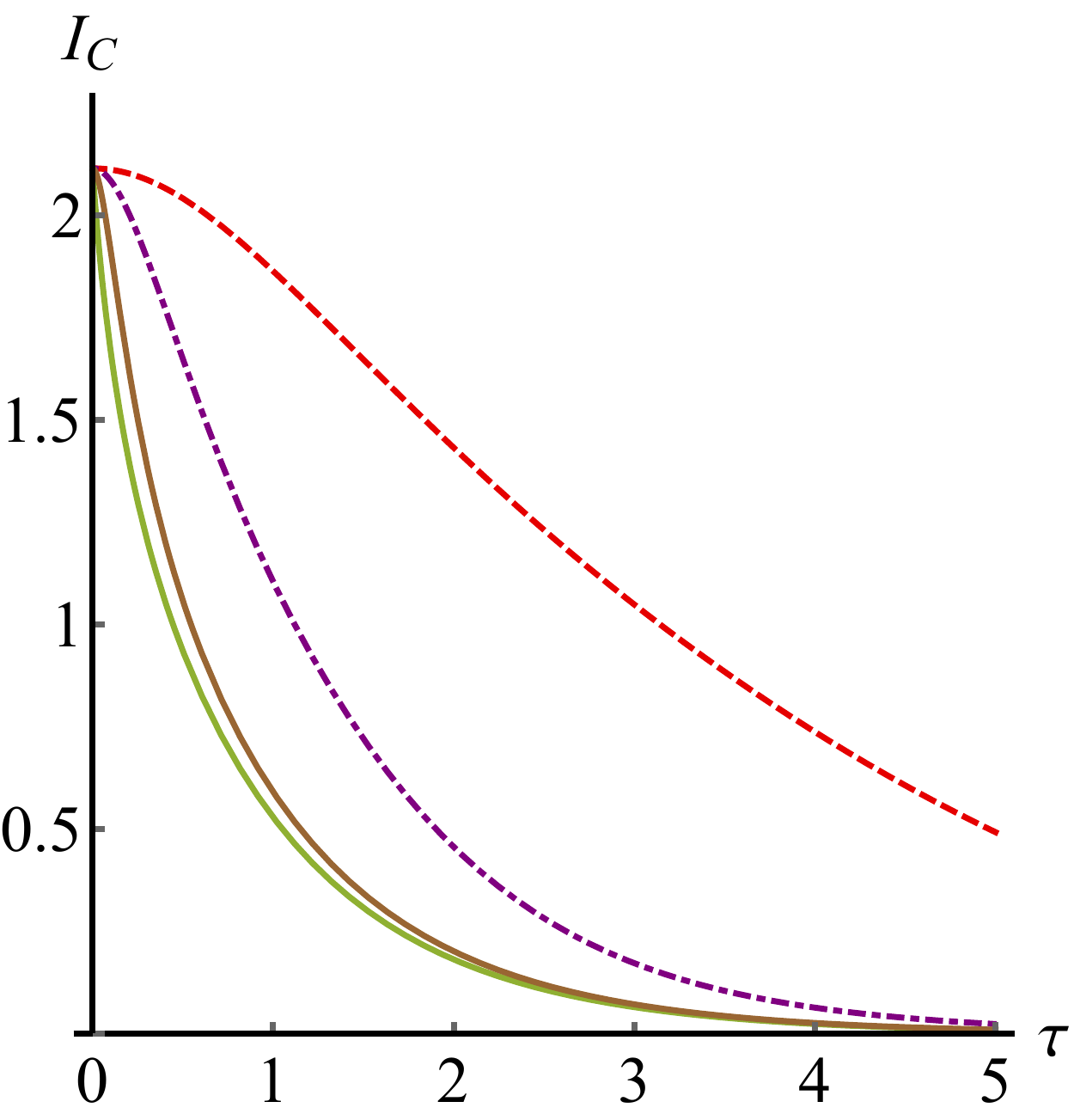}
\includegraphics[width=0.45\columnwidth]{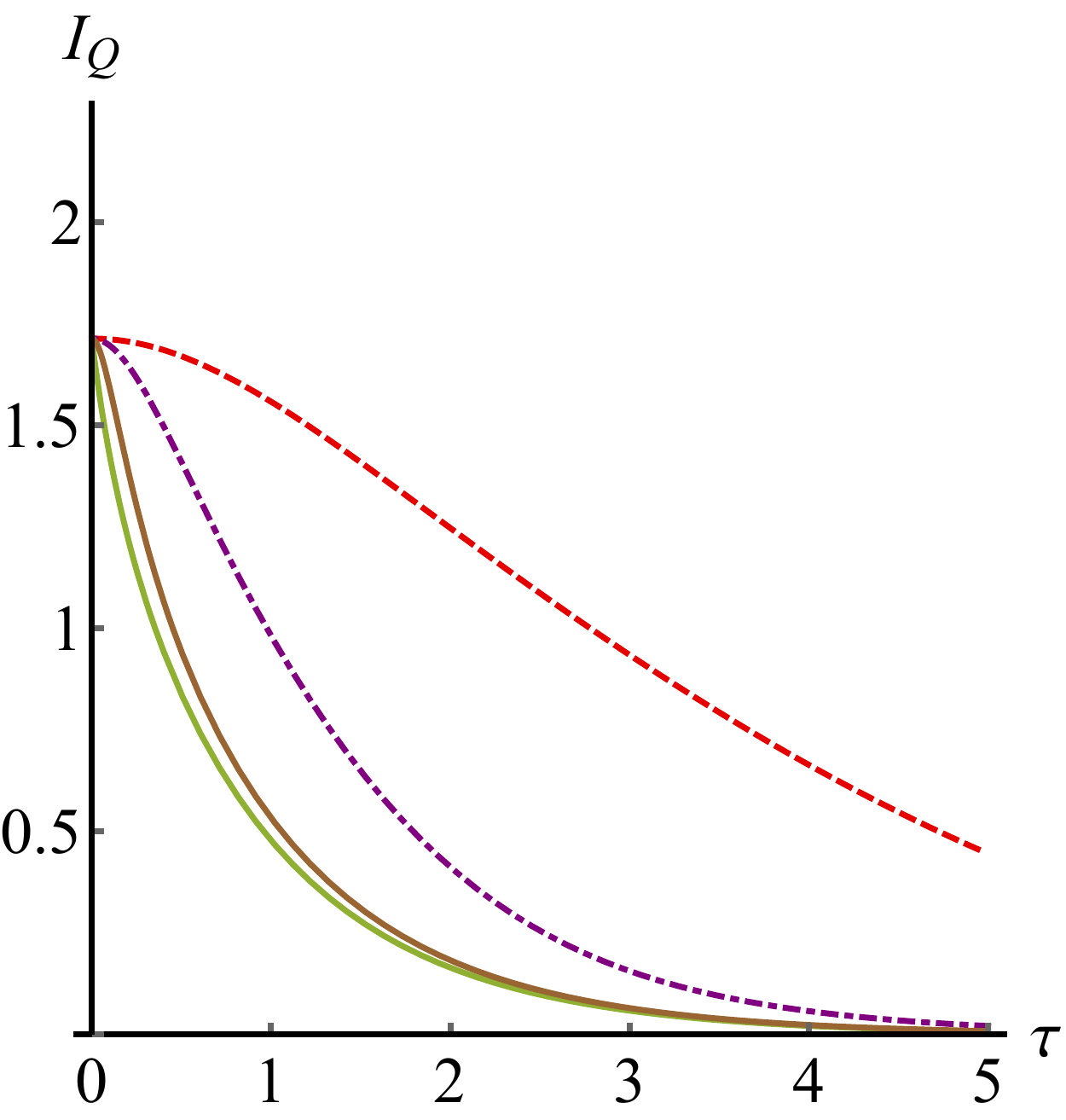}\\
\vspace{0.5cm}
\includegraphics[width=0.7\columnwidth]{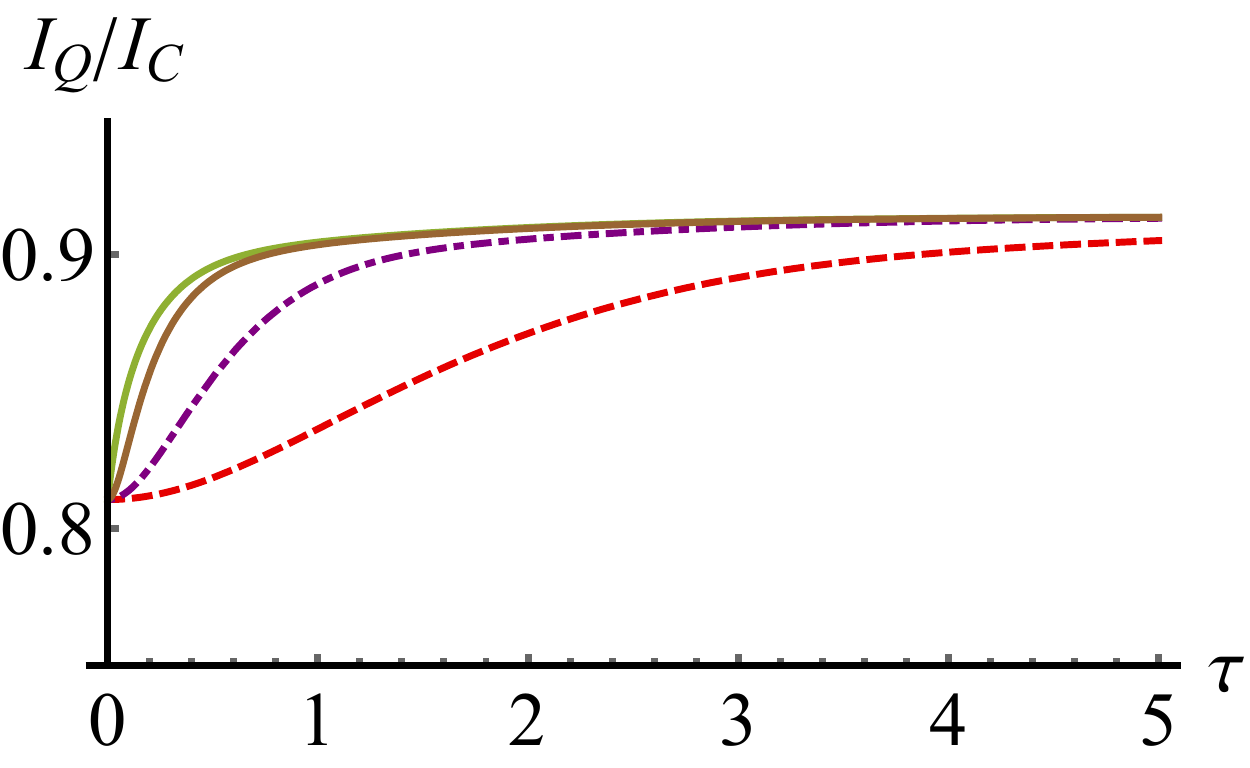}
\end{center}
\vspace{-0.5cm}
\caption{\label{f:stocIDIQ} (Color Online) Phase 
communication channels in the presence of dynamical phase 
diffusion. The upper panels show the mutual informations $I_\sid$ (left) 
and $I_\sq$ (right) as function of $\tau = \Gamma t$ for different values of 
the correlation time $\tau_\sE$ of the environment.
From bottom to top, $\tau_\sE=0.1$ (solid brown), $1$ (dotdashed 
purple) , $10$ (dashed red). The lower solid green curve is the mutual 
information in the static case.
The other parameters read as follows: $N=20$, $\lambda=1$, 
$\bar n = 3$. The lower panel shows the ratio $I_\sq/I_\sid$
as a function of $\tau$ for the same values of $\tau_\sE$ and
of the other parameters. The color code is the same of the upper 
panels.}
\end{figure}
In the following, we consider a generalized phase diffusion model 
corresponding to the quantum map
\begin{equation}
\label{eq:channel}
\rho(\tau) = \int_{-\infty}^{\infty}\!
\frac{d\phi}{\sqrt{2 \pi \sigma(\tau)}}\, 
e^{-\frac{\phi^2}{2 \sigma(\tau)}}\,
U(\phi) \rho(0) U^{\dagger} (\phi),
\end{equation}
where $\sigma(\tau)$ is a time-dependent variance, summarizing the
dynamical properties of the environment, and, for convenience, we
still use the rescaled time $\tau = \Gamma t$. The static 
environment of the previous Section is recovered for $\sigma(\tau)= 
\tau$. The quantum map (\ref{eq:channel}) turns
the input state $\rho_k$ into a statistical mixtures of states 
with a time-dependent Gaussian distribution of the phase around 
the original phase $\phi_k$. The time dependence of $\sigma(\tau)$ 
is linked to the correlation function of the underlying stochastic 
noise as follows
\begin{equation} 
\sigma(\tau) = \int_0^t\! ds_1  \int_0^t\!  ds_2
\cos  [\delta (s_1 -s_2)]\, K(s_1, s_2) \,,
\end{equation}
\JT{where $K(s_1,s_2)$ is the correlation function of the specific 
CSF chosen to describe the noise
and $\delta=(\omega_0 - \omega) /\Gamma$ is the rescaled detuning
between the carrier frequency $\omega_0$ and the central frequency
of the environmental spectrum $\omega$.}
In this paper, we focus on the noise generated by the 
Ornstein- Uhlenbeck process with a Lorentzian spectrum and 
correlation function $K(\tau_1,\tau_2) = \frac12 
\lambda\, \tau_\sE^{-1} \, \exp[-|\tau_1-\tau_2| /\tau_\sE]\,$,
where $\tau_\sE$ is the characteristic time of the environment and \JT{$\lambda$ is the dynamical phase noise factor, rescaled with $\Gamma$.}
In this case, $\sigma(\tau)$ is given by
\begin{align}
\sigma(\tau) =& \frac{\lambda}{[ 1 + (\delta\, \tau_\sE)^2]^2}
\Big\{\tau- \tau_\sE + (\delta\, \tau_\sE)^2 (\tau+ \tau_\sE)\notag \\ &+ \tau_\sE\, 
e^{-\tau/\tau_\sE}\left[(1-(\delta\, \tau_\sE)^2)\cos \delta\, \tau - 
2 \delta\, \tau_\sE \sin \delta\, \tau\right]\!\Big\}.
\end{align}
In the Markovian limit $ \tau_\sE \ll \tau$, the latter may be re-written as 
\begin{equation}
\label{eq:sigmalimit}
\sigma(\tau)  \simeq \lambda \tau 
\end{equation}
whereas, in the presence of highly correlated environments  $\tau_\sE \gg \tau$ , it becomes
\begin{equation}
\label{eq:sigmalimit2}
\sigma(\tau)  \simeq \frac12 \lambda \tau^2/\tau_\sE.
\end{equation}
Equation (\ref{eq:sigmalimit}) confirms that the quantum map 
(\ref{eq:channel}) is the solution of the Markovian master equation 
for a static phase-diffusing environment, upon setting $\lambda = 1$. 
In this case the environment is characterized by a very short correlation 
and the stochastic field describes a Markovian interaction. The
corresponding dynamics of mutual information 
approaches that illustrated in Fig. \ref{f:mutInf1}.
\par 
If the environment shows non-zero correlation time 
the dynamics of mutual information may be dramatically altered, 
showing either a different decay rate or the appearance of oscillations.
In the following we first analyze the case of a 
{\em resonant enviroment} with zero detuning $\delta=0$ and 
then focus attention to nonresonant situations.
In both cases, the probabilities $q_k(s)$, $k=C,Q$ are still 
given by Eqs. (\ref{eq:qID}) and (\ref{eq:MUQ}) with the replacement
$$
\exp \left(-\frac12 d^2 \tau \right) \longrightarrow
\exp \left[-\frac12 d^2 \sigma(\tau)\right]\,.
$$
\par
Let us start with the case of a resonant environment ($\delta = 0$).
Under such condition, $\sigma(t)$ reduces to
\begin{equation}
\sigma(\tau) = \lambda \left[\tau - \tau_\sE (1-e^{-\tau/\tau_\sE})\right]
\end{equation}
and the channel appears to be more robust against the effects of noise, 
at least for a short time dynamics, compared to the static case.
In order to illustrate this feature, in Fig. \ref{f:stocIDIQ}, 
we show the mutual informations $I_\sid$ and $I_\sq$ as a function 
of $\tau$ for different values of $\tau_\sE$. As it is apparent 
from the plot, the presence of a non-zero correlation time of the environment 
$\tau_\sE$ better preserves mutual information against phase diffusion for 
both ideal and $Q$ receiver. As it happens in the static case the 
mutual information vanishes with time. However, a time-correlated 
environment allows a ``concave dynamics'' of the mutual information, which 
lasts longer, the higher is the correlation time. This behaviour is due 
to the transition from linear to quadratic behaviour of $\sigma(\tau)$, see Eq.
(\ref{eq:sigmalimit2}), which may be observed for increasing $\tau_\sE$.
We also show the mutual information for the static case (solid green
line) for comparison.
In the lower panel of the same Figure we report the ratio $I_\sq/I_\sid$
as a function of $\tau$. Upon comparing this plot with 
the lower panel of Fig. \ref{f:mutInf1} we conclude that dynamical
noise is more detrimental for $Q$ receivers than for ideal ones.
\begin{figure}[tb!]
\vspace{0.5cm}
\begin{center}
\includegraphics[width=0.45 \columnwidth]{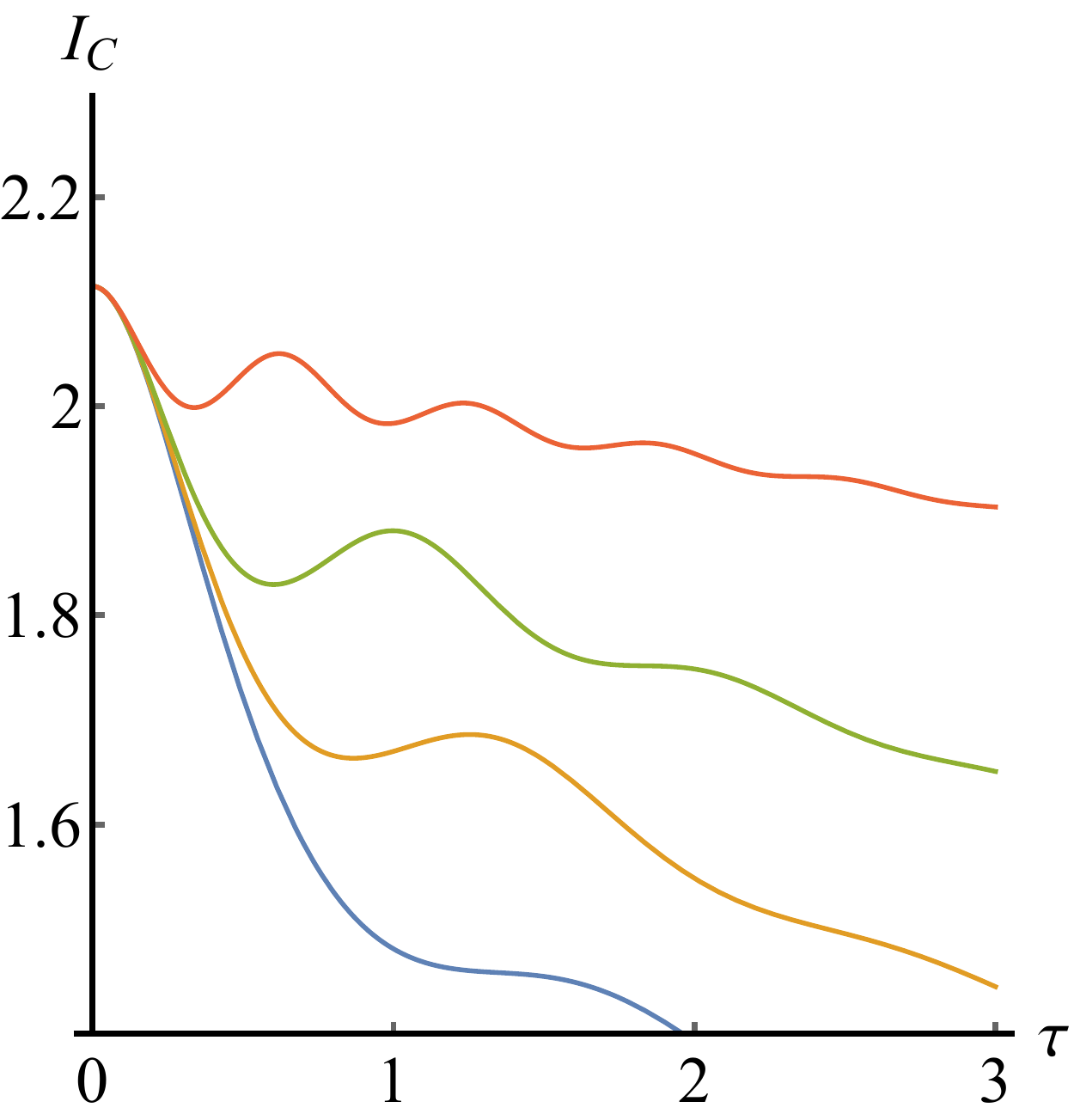}
\includegraphics[width=0.45 \columnwidth]{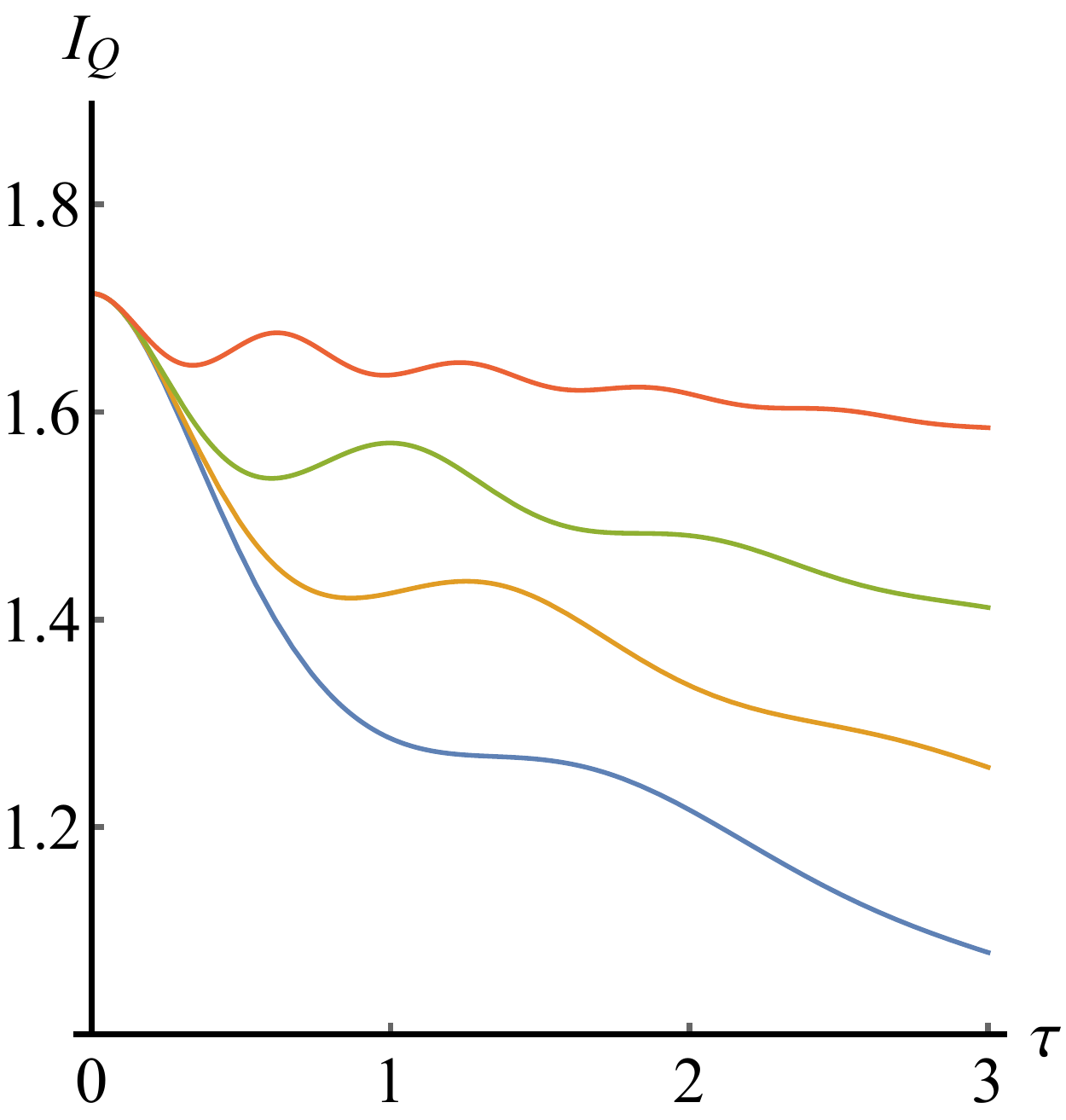}\\
\vspace{0.5cm}
\includegraphics[width=0.45 \columnwidth]{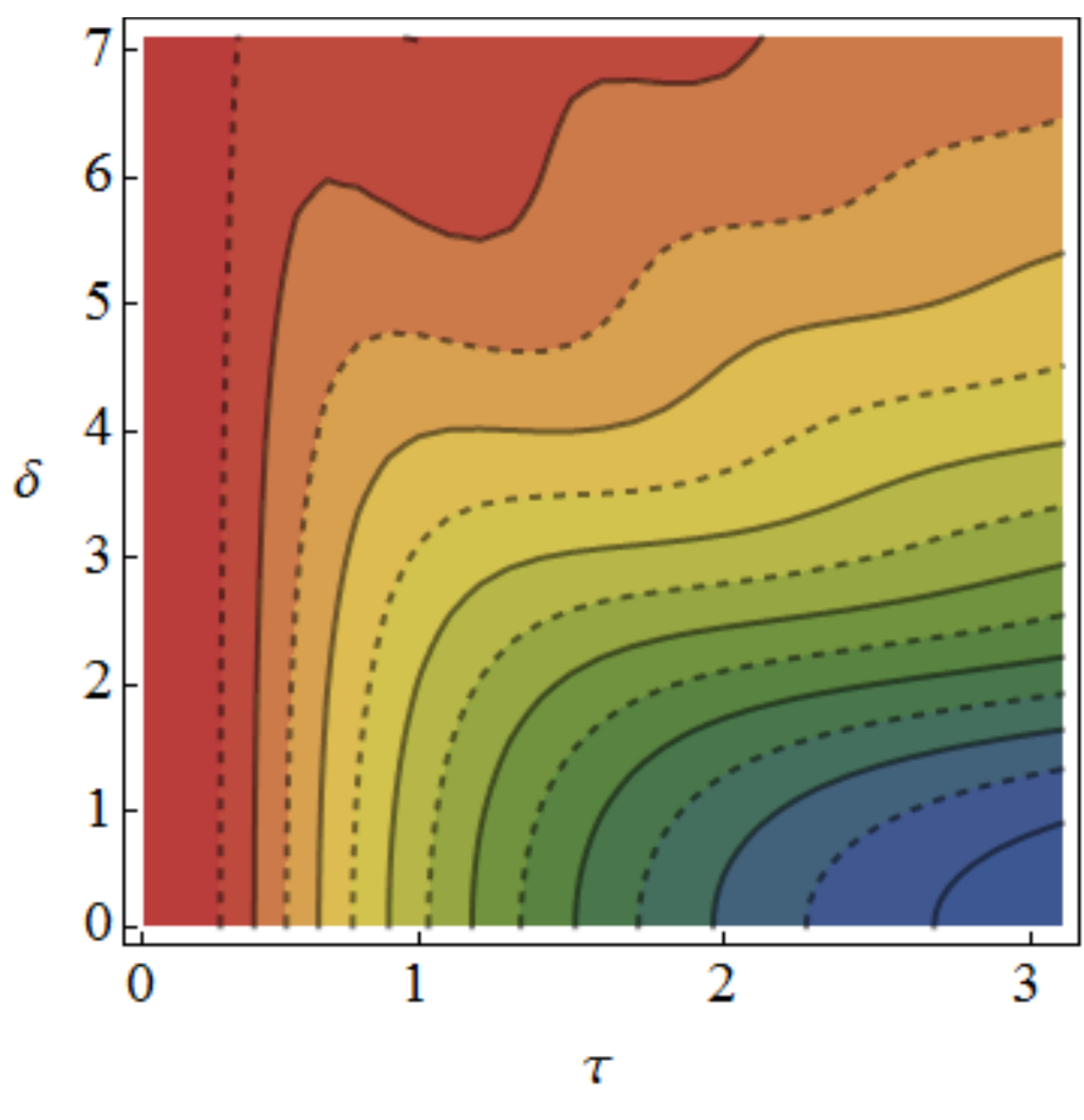}
\includegraphics[width=0.45 \columnwidth]{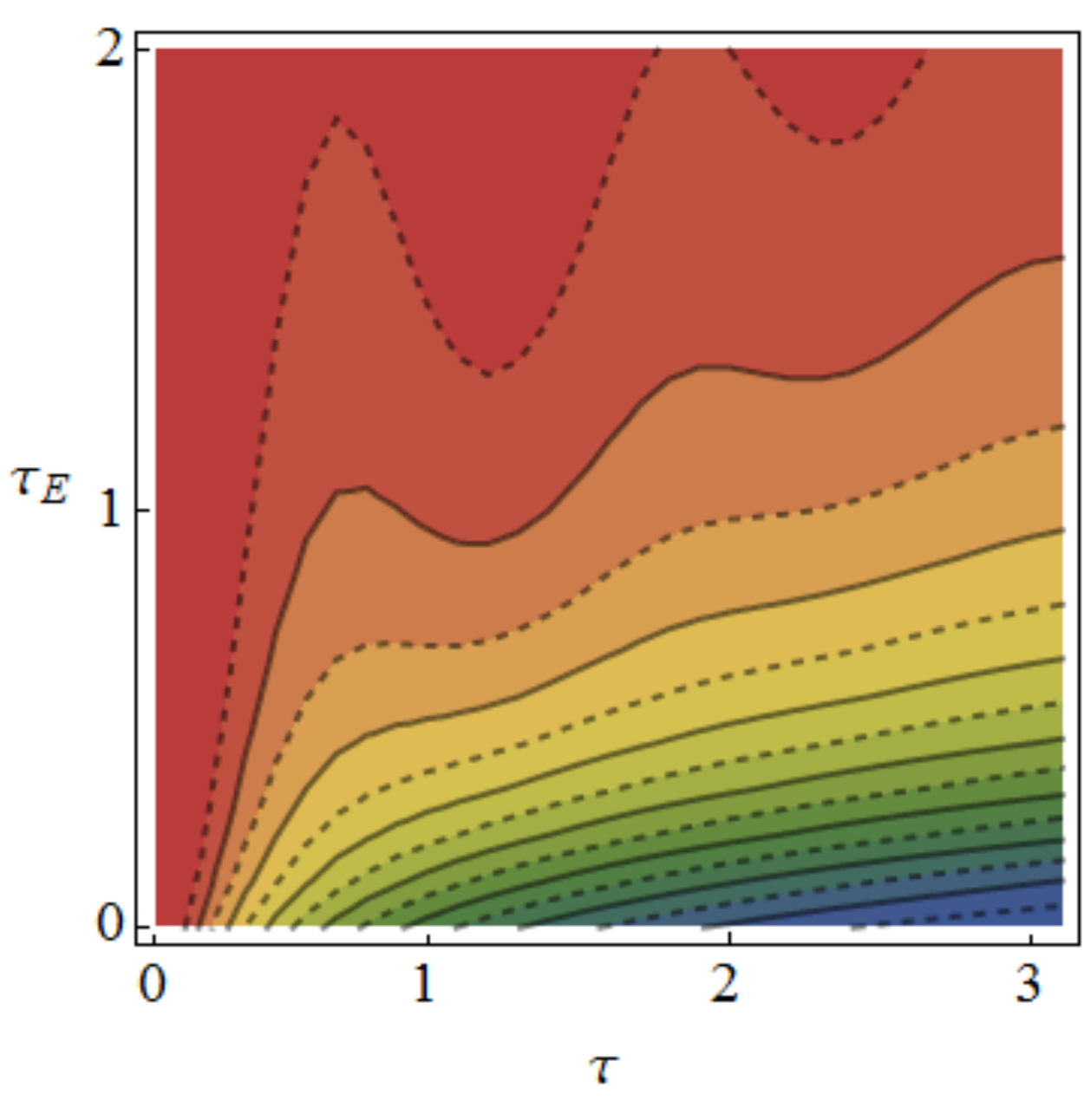}
\end{center}
\vspace{-0.5cm}
\caption{\label{f:figdet} (Color online) Phase communication channels 
in the presence of dynamical phase diffusion. The upper panels show the
mutual information $I_\sid$ (left) and $I_\sq$ (right) as a function
of $\tau = \Gamma t$ for different values of
detuning. From top to bottom $\delta = 10$ (red), $\delta=6$ (green), 
$\delta=4.5$ (orange), and  $\delta=3.5$ (blue). The other parameters
are given by $N=20$, $\lambda=1, \bar n = 3, \tau_\sE = 1.$ 
The lower left panel shows the contour plots of $I_\sid$ as a function of 
$\tau$ and detuning $\delta$ for $N=20$, $\lambda=1$, $\bar n = 3$, and 
$t_\sE = 1$. The right panel contains the contour plots $I_\sq$ as a function 
of $\tau$ and $t_\sE$ for $N=20$, $\lambda=1$, $\bar n = 3$, 
and $\delta= 5.5$.}
\end{figure}
\par
Let us now analyze the effects of detuning between 
the frequency of the information carrier and the central 
frequency of the CSF. As it is possible to see from 
the upper panels of Fig. \ref{f:figdet}, the dynamics of 
the mutual information is strongly affected by the detuning
for both kind of receivers. On the one hand, the detuning 
contributes to the significative slowdown of the damping of 
mutual  information and, on the other hand, it is responsible 
for the appearance of revivals of mutual information, which 
can be interpreted as
a sign of a backflow of information caused by the 
non-Markovian effect of the detuned dynamical map.
Yet, the contourplots of mutual information, shown in 
the lower panels of the same Figure, reveal that the presence of
revivals is also related to the correlation time
of the environment. In the left panel, we show
that for fixed correlation time of the environment $\tau_\sE = 1$ 
revivals appear only for particular values of detuning $\delta$.
In the right panel, we show that for fixed value of detuning
$\delta = 5.5$ revivals appear beyond a threshold value of
the correlation time of the environment.
\section{Conclusions}
\label{s:out}
We have analyzed quantum phase communication channels based on 
phase modulation of coherent states and addressed their performances 
in presence of static and dynamical phase diffusion by evaluating 
the mutual information for ideal and realistic phase receivers.  
In terms of performance, our results show that phase communication 
channels are robust, especially for large alphabets in the low energy 
regime, and that their performances are comparable to those of 
coherent channels in the presence of loss.  
\par
In the presence of dynamical (non-Markovian) phase diffusion, 
phase channels become more robust, the mutual information being 
preserved by the time correlations of the environment. 
When the noise spectrum is detuned with respect to the information 
carrier, revivals of mutual information  also appear.
\par
Our results illustrate the potential applications of phase-keyed
$M$-ary channels and may be also of interest in other schemes where the 
information is coded on phase shifts as, for example, in
interferometric high-sensitivity measurements. 
\section*{Acknowledgments}
This work has been supported by MIUR through the FIRB project
``LiCHIS'' (grant RBFR10YQ3H), by EU through the Collaborative 
Projects and QuProCS (Grant 
Agreement 641277) and by UniMI through the H2020 Transition 
Grant 14-6-3008000-625. This paper is dedicated to the memory of 
Gabriele Corbelli.

\end{document}